\documentclass[aps, notitlepage, onecolumn, pre, superscriptaddress, floatfix, nofootinbib, showkeys]{revtex4-2}

\usepackage{pgfplots}
\usetikzlibrary{intersections}
\pgfplotsset{compat=1.17}
\usepackage{amsfonts,amsmath,amssymb}
\usepackage{mathtools}   
\usepackage[normalem]{ulem}
\usepackage{hyperref}
\usepackage[flushleft]{threeparttable}
\usepackage{booktabs, makecell}
\usepackage{xcolor}
\usepackage[ruled]{algorithm2e}
\usepackage{colortbl} 
\usepackage{placeins} 
\usepackage [english]{babel}
\usepackage [autostyle, english = american]{csquotes}

\usepackage{xcolor}
\definecolor{goodblue}{RGB}{0, 91, 187}
\hypersetup{
  colorlinks=true,
  allcolors=goodblue,
  urlcolor=goodblue,
  citecolor=goodblue,
  pdfborder={0 0 0},
  breaklinks=true,
}

\MakeOuterQuote{"}

\begin{document}

\title{Higher-order interactions at scientific conferences influence team formation}

\author{Emma Zajdela}
\email{emma.zajdela@princeton.edu}
\affiliation{High Meadows Environmental Institute, Princeton University, Princeton, NJ, USA}
\affiliation{Santa Fe Institute, Santa Fe, NM, USA}

\author{Nicholas W. Landry}
\email{nicholas.landry@virginia.edu}
\affiliation{Department of Biology, University of Virginia, Charlottesville, VA, USA}
\affiliation{School of Data Science, University of Virginia, Charlottesville, VA, USA}
\affiliation{Vermont Complex Systems Institute, University of Vermont, Burlington, VT, USA}

\begin{abstract}
Cooperation enables teams to solve complex problems that one individual alone cannot address. In science, collaborative teams have become the predominant way through which progress is achieved. These scientific collaborations arise though various mechanisms, among which interactions at conferences. The Scialog conferences, which comprise a series of small, interdisciplinary scientific workshops held over several years, are an ideal laboratory to study the network mechanisms leading to team formation. Building on existing work studying team formation from a pairwise perspective, we present a higher-order network perspective generalizing this framework. We provide a formalization for the notion of group interaction over time by defining a taxonomy of synchronous and asynchronous group interactions. We apply this framework to the Scialog case study using a stepwise selection logistic model and find evidence that all interaction types described in our taxonomy are highly significant for team formation. This higher-order network perspective provides a new framework for the study of collective behavior and group formation.
\end{abstract}

\keywords{higher-order networks; Science of Science; team formation; collective intelligence} 
\maketitle




\section{Background}

Many current and future challenges require individuals to collaborate across disciplinary boundaries \cite{meluso2023multidisplinary}. Groups of individuals can assemble into teams and solve problems that one individual alone cannot, thereby demonstrating collective intelligence \cite{leonard2022}. These collaborative teams play an important role in generating knowledge, innovation, and discoveries across many facets of society. As the importance of cooperation in teams has been demonstrated, there has been a recent increase in the study of teams in general, leading to the development of the field of Science of Team Science (see review papers including \cite{stokols2008,bozeman2014,liu2020}). 

The Intelligence Community requires teamwork and collaboration across stakeholders with different backgrounds, skills, and experience, requiring coordination within and across teams. These collaborations are faced with challenges due to the nature of the Intelligence Community's work, including the need for secrecy and cross-disciplinary teams, as well as communication challenges that arise when coordinating individuals from institutions with their own norms and culture \cite{hackman2011,vogel2019}. Several of the challenges faced by teams in the intelligence community are shared by the scientific community, including difficulties in communication when working on problems that require individuals from different backgrounds, disciplines, and institutions to collaborate. Despite these challenges, across nearly all scientific disciplines, research is increasingly performed in teams. In science and engineering, the fraction of publications written in teams has increased from around 50\% in 1955 to over 80\% by 2000 \cite{wuchty2007}. As more scientific knowledge is produced in teams, the papers authored by these teams are more cited than solo-authored work \cite{wuchty2007}. 

Several studies have investigated the assembly of teams in science \cite{guimera2005,Lungeanu2014}. In 2015, a survey showed that for collaborators who were not geographically co-located, one out of six met at conferences \cite{freeman2014}. When considering the role of events like scientific conferences, research has demonstrated the role of prior-knowledge \cite{Lungeanu2014} and interaction \cite{Zajdela2022,zajdela2025,lane2021,boudreau2017} for team formation. These studies focus on predicting the pairwise effects between participants at the conference, despite the fact that scientific collaborations and the groups in which scientists interact at conferences are often larger than two.

Higher-order interactions, which represent interactions of arbitrary size between individuals, more naturally model these systems and can offer insights inaccessible to pairwise representations.
Higher-order networks, or \textit{hypergraphs}, are the collection of these higher-order interactions and naturally encode the scale of interaction~\cite{landry2024filtering}.
Even when measuring properties of a collaboration network such as its community structure, assortativity, and the relative importance of its constituent nodes, modeling these networks in their native representations can change the conclusions that one comes to.
For example, when examining scientific collaborations\cite{Patania2017}, higher-order network analysis indicates that while the sizes of collaborations that researchers engage in is field dependent, that the number of collaborations in which they engage remains relatively consistent.
Likewise, hypergraph analysis preserves group-level interactions, allowing for analysis of higher-order collaboration motifs~\cite{juul2024hypergraph}.
Lastly, higher-order networks provide an essential framework for studying the temporal evolution of group structure~\cite{iacopini2024temporal}.
Pairwise network representations are unable to disambiguate between groups of different size, and therefore cannot be used to how groups split, merge, and aggregate over time.

In order to study the impact of interaction on group outcomes, including team formation, we first need to provide a definition of interaction over time for groups of more than two individuals.
Given known effects in social networks including triadic closure, this chapter also seeks to identify the extent to which higher-order interaction adds significant information to group outcomes.
Therefore, we extend the pairwise methods historically used to analyze these systems by defining a taxonomy of synchronous and asynchronous group interactions.
We use this taxonomy to identify types of interactions occurring through sub-groups that may contribute to group outcomes.
We apply these notions to a case study comprising four scientific conferences from the Scialog dataset, containing detailed, longitudinal information about team formation at interdisciplinary scientific conferences known as "the Scialogs" \cite{wiener2019}.
We find evidence that all types of interactions described in our group interaction taxonomy impact team formation.
This framework can be applied to other cases of interaction in social networks over time in contexts such as teams in the Intelligence Community, business, and medical settings, as well as other group interactions such as human migration or animal behavior.

\section{Methods}

In this section, we describe the use of higher-order network analysis to predict group collaboration.
We start by defining the interaction hypergraph $H_i = (V_i, E_i)$ and the collaboration hypergraph $H_c=(V_c, E_c)$.
We assume that $V_i=V_c=V$, and let $N = |V|$ denote the size of each hypergraph. 
For each group session $e\in E_i$, we track the start time $t^{(i)}_e$ and end time $t^{(f)}_e$.
Consistent with previous work defining interaction in the pairwise case \cite{Zajdela2022}, we make the following three assumptions: (1) interaction is symmetric, (2) interaction is proportional to the time participants spend listening to one another, and (3) participants speak an equal amount of time in a given group session.
These assumptions specify an interaction function $I_e =(t^{(f)}_e-t^{(i)}_e)/|e|$ and imply that interaction in smaller groups is more intense than larger groups. 

Our goal is to predict collaboration groups of size 3; i.e., we wish to understand the interaction patterns present in $H_i$ that give rise to a collaboration $e\in E_c$, where $|e|=3$.
We do this by introducing a group interaction taxonomy that aims to uncover the role of pairwise and higher-order group formation mechanisms.
This paradigm is described in more detail below.

\begin{figure}[bp] 
    \centering
    \includegraphics[width=\linewidth]{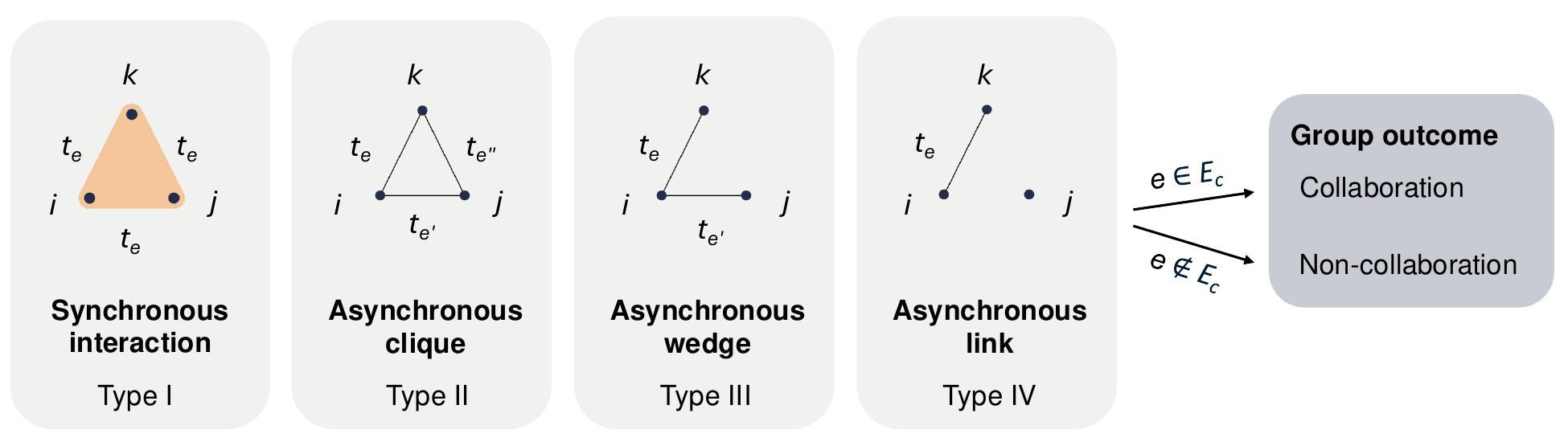}
    \caption{For groups of three, this figure shows the possible types of interaction that can occur. Type I: the three participants have interacted in the same group at least once (synchronous), type II: the three individuals have interacted with each other at different points in the conference but never at the same time, type III: two out of three possible pairs involving the three individuals have interacted, but never at the same time, and type IV: one out of the three possible pairs involving the three individuals have interacted.}
    \label{fig:def_interaction}
\end{figure}

Given a group $\{i, j, k\}$, we classify groups in $H_i$ according to the following taxonomy:

\begin{enumerate}
    \item \textbf{Synchronous interaction (Type I):} The group  $\{i, j, k\}$ has a synchronous interaction in breakout session $e$ if $\{i, j, k\} \subseteq e$.
    
    \item \textbf{Asynchronous interaction:} An asynchronous interaction occurs in group $e$ if $\{i, j\}\subseteq e$, $\{i, k\}\subseteq e$, or $\{j, k\}\subseteq e$ while $\{i, j, k\}\not\subseteq e$.
    Aggregated over all groups $e\in E_i$, asynchronous interactions can occur between three, two, or one of the pairs in the group, corresponding to types II, III, and IV respectively as seen in Fig.~\ref{fig:def_interaction}
    \begin{enumerate}
        \item \textbf{Asynchronous clique (Type II)}: This measures the minimum amount of time that $\{i, j\}$, $\{i, k\}$, and $\{j, k\}$ interacted separately over the course of the conference.
        \item \textbf{Asynchronous wedge (Type III)}: This measures the minimum amount of time that two of the three possible pairs $\{i, j\}$, $\{i, k\}$, and $\{j, k\}$ interacted separately over the course of the conference.
        \item \textbf{Asynchronous link (Type IV)}: This measures the minimum amount of time that a single pair from $\{i, j\}$, $\{i, k\}$, and $\{j, k\}$ interacted separately over the course of the conference.
    \end{enumerate}
\end{enumerate}

We operationalize this taxonomy using the following algorithm to compute the type I-IV interaction times corresponding to a collaboration (or non-collaboration) group $g$.
First, the algorithm searches over all possible sessions $e\in E_i$ and looks for all subgroups of group $\tilde{g}\subseteq g$ in order of decreasing size.
If $\tilde{g} \in e$, the algorithm increments the interaction time by $I_e$ and keeps iterating over all remaining groups.
The result of this first step is a map $\theta: g\to \mathbb{R}$ between subsets $\tilde{g}$ and interaction times.
Once this map is completed, the type I interaction time is simply $\theta(g)$.
To compute the types II-IV interaction times, we do the following: (1) construct a vector $\mathbf{a}=[\theta(\tilde{g})\mid \tilde{g}\in g \ s.t. \ |\tilde{g}| = 2]^T$, (2) let $\delta = |\{a_i\in \mathbf{a}\mid a_i >0|$, (3) set $\tilde{t}(2, \delta) = \min \mathbf{a}$ where $\tilde{t}: (i, j)\to \mathbb{R}$, (4) let $\mathbf{a} = \mathbf{a} - \tilde{t}(2, \delta)$ and $\mathbf{a} = [a_i \in \mathbf{a} \mid a_i > 0]^T$\;, and (5) repeat steps 2-4 until $\mathbf{a}=\emptyset$.
Lastly, we translate the map $\tilde{t}$ to the interaction types define in Fig.~\ref{fig:def_interaction}.
The type II interaction time is $\tilde{t}(2, 3)$, the type III interaction time is $\tilde{t}(2, 2)$, and the type IV interaction time is $\tilde{t}(2, 1)$.

The taxonomy described in Fig.~\ref{fig:interactions} isolates the effects of distinct mechanisms thought to influence group formation.
The type I "synchronous" interactions illustrate the role of higher-order effects on group formation.
This mechanism requires the individuals to interact in the same group and cannot occur when individuals meet individually over the course of a conference.
Likewise, the type II "asynchronous clique" interaction posits that all individuals must have met, just not in the same context.
Similarly, if type III "asynchronous wedge" interactions are predictive of collaboration groups, this indicates that triadic closure is at play.
Lastly, type IV "asynchronous link" interactions describe group formation through random aggregation, where individuals who have never formally interacted with a group of two may join it nonetheless.
Simpliciality, defined in Ref.~\cite{landry2024simpliciality}, measures the inclusion structure of higher-order networks and further study is needed to link the interaction types described above with a functional (weighted) notion of simpliciality.

\section{Case study: Scialog conferences}

\begin{figure}[ht]
    \centering
    \includegraphics[width=\linewidth]{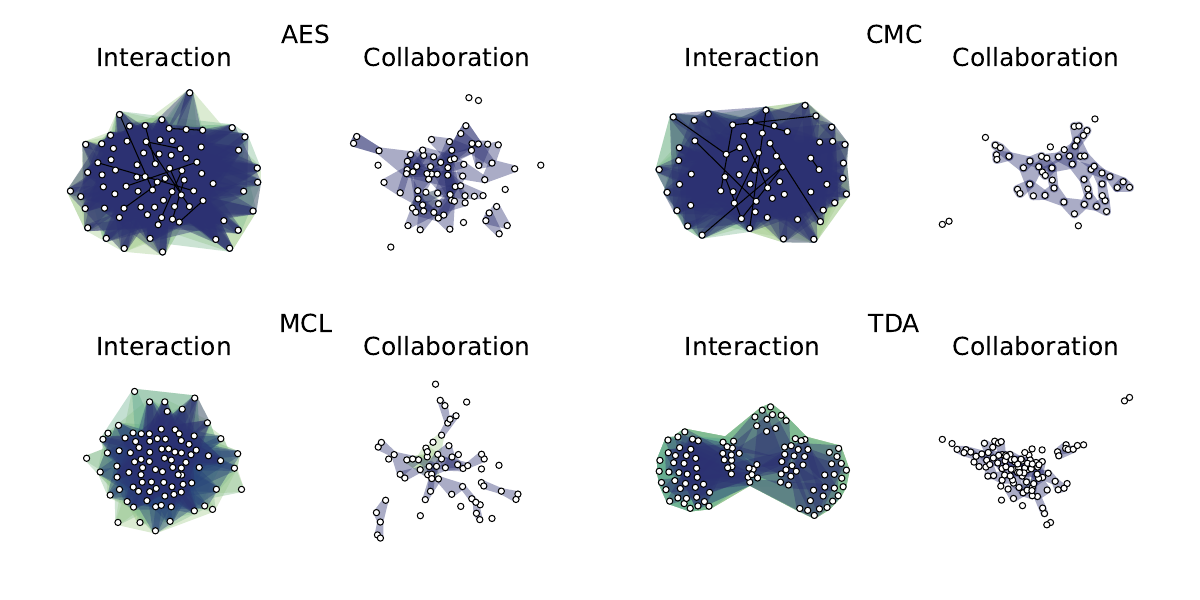}
    \caption{An illustration of the network structure of each conference series. Each network is aggregated over all conferences in that conference series. Visualized with XGI~\cite{landry2023xgi}.}
    \label{fig:network-viz}
\end{figure}

We illustrate our higher-order framework with a case study utilizing the Scialog conference \cite{wiener2019} datasets.
The data used in our case study is from four interdisciplinary conference series as part of the Scialog program conducted by the Research Corporation for Science Advancement: the "Advanced Energy Storage" (AES) conference run in years 2017-2019, the "Chemical Machinery of the Cell" (CMC) conference run in years 2018-2021, the "Molecules Come to Life" (MCL) conference run in years 2015-2017, and the "Time Domain Astrophysics" (TDA) conference run in years 2015-2016 and 2018-2019.
These conferences take place in-person over the course of three days, with each conference containing fewer than 70 participants who are primarily early-career faculty from the US and Canada, as well as some senior experts who serve as keynote speakers and facilitators.
The structure of the conference is designed to maximize interaction among participants, who are algorithmically assigned to medium group discussions and small group discussions (more information about the group assignment algorithm is available in \cite{zajdela2025}).
At the end of the conference, participants self-assemble into teams of $2-4$ and submit proposals of which $17-33\%$ receive funding. 

The conference interaction data presented here comprises two different types of interaction data: (1) records of attendees participating in medium group discussions (10 participants or more) and small group discussions (4 participants or fewer) and (2) records of the collaborations that participants eventually formed.
The interaction and collaboration data for each conference are visualized in Figure~\ref{fig:network-viz} and a brief summary of selected network statistics for each dataset is presented in Table~\ref{tab:network_stats}.

\begin{table}[b]
\centering
\begin{tabular}{lccccccc}
Dataset&\null\quad\null & $|V|$ & $|E|$ & $\langle k\rangle$ & $\langle k_{pairwise}\rangle$ &$\langle s\rangle$ & Unique interaction sizes\\
\hline\\
\textbf{AES 2017}&&&&&\\
interaction && 60 & 104 & 8.0 & 44.0 & 4.62 & 3, 10\\
collaboration && 56 & 35 & 1.59 & 2.64 & 2.54 & 2, 3, 4\\[0.1in]
\textbf{CMC 2018}&&&&&\\
interaction && 50 & 88 & 8.0 & 43.84 & 4.55 & 2, 3, 10\\
collaboration && 45 & 24 & 1.42 & 2.49 & 2.67 & 2, 3\\[0.1in]
\textbf{MCL 2015}&&&&&\\
interaction && 64 & 54 & 5.33 & 40.12 & 6.31 & 4, 12, 13\\
collaboration && 40 & 20 & 1.3 & 2.3 & 2.6 & 2, 3, 4\\[0.1in]
\textbf{TDA 2015}&&&&&\\
interaction && 49 & 72 & 8.0 & 46.53 & 5.44 & 3, 4, 9, 10\\
collaboration && 45 & 29 & 1.53 & 2.27 & 2.38 & 2, 3\\[0.1in]
\hline\\
\end{tabular}
\caption{The network properties of the Scialog datasets. From left to right, the columns indicate the number of nodes, the number of interactions, the average higher-order degree, the average pairwise degree, the average interaction size, and the unique interaction sizes at that conference and for that interaction type (interaction or collaboration). For each conference, "interaction" is a subset of the conference dataset comprising pre-collaboration interactions through the large and small discussion groups to which participants are assigned. Likewise, "collaboration" denotes the subset of the conference dataset comprising the collaborations that participants formed.}
\label{tab:network_stats}
\end{table}

We can see that the higher-order degree provides little to no information on the number of groups a participant joins.
In addition, while interaction groups can be quite large due to the medium group, the resulting collaboration groups are predominantly of sizes two and three.
These group sizes are due to the assignments to discussion groups and the rules for the proposal submissions, which typically require participants to be in teams of $2-3$ members. 

\begin{figure}
    \centering
    \includegraphics[width=0.8\linewidth]{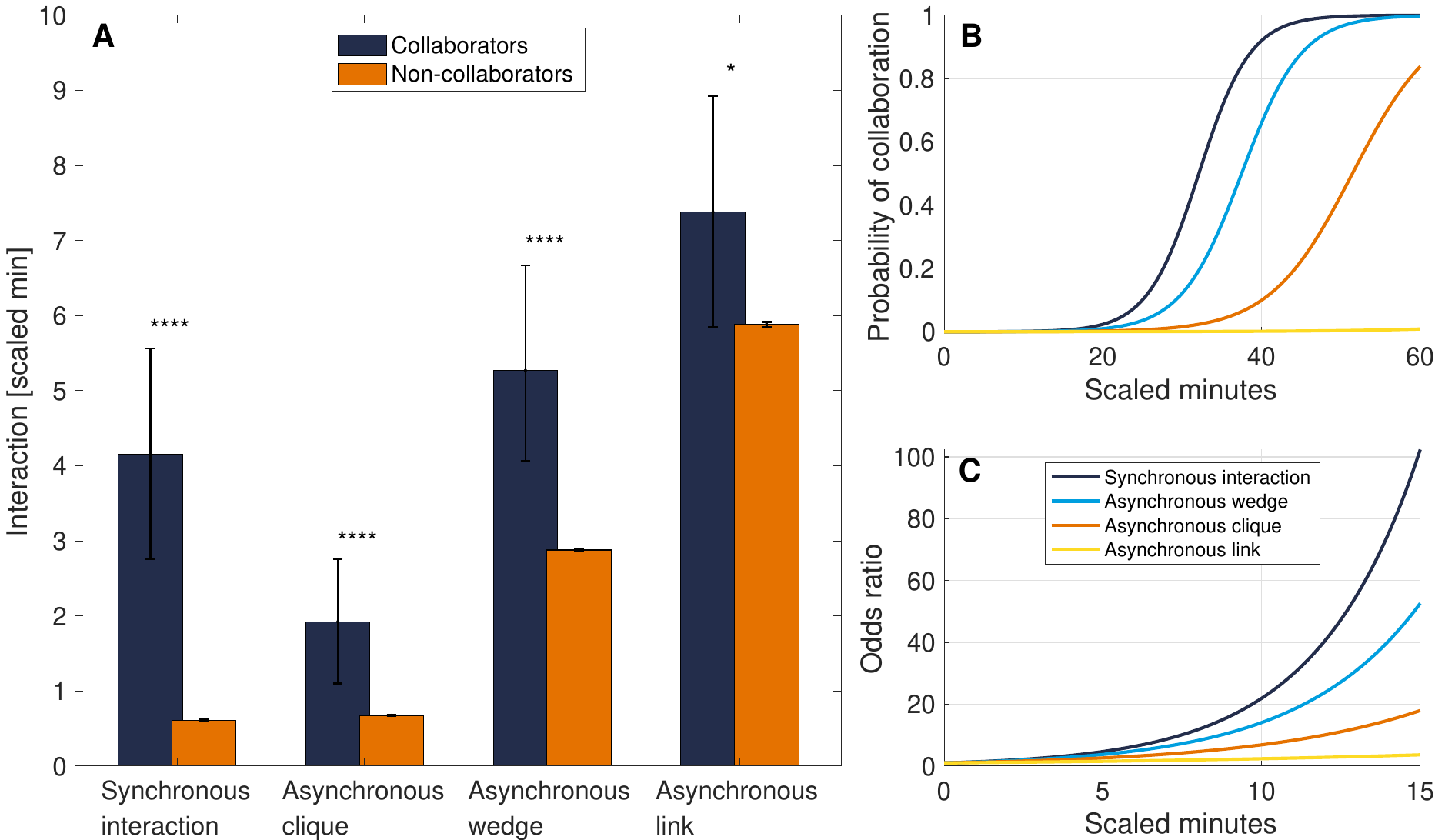}
    \caption{\textbf{Interaction and collaboration: } Panel A shows mean total scaled interaction time [scaled minutes] for people who ended up collaborating (left bars) or not (right bars) in a group of three. Data is aggregated across all four conferences and is shown for the four types of interaction defined in the main text. Error bars show mean values of the bootstrapped data with 95\% CI. P-values of the Mann–Whitney U test: synchronous interaction, $1.04\times10^{-21}$; asynchronous clique, $3.3\times10^{-2}$; asynchronous wedge, $6.7\times10^{-6}$; asynchronous link, $3.1\times10^{-5}$. Panel B and C are a visualization of the logistic model results. Panel B shows the change in probability of collaboration as scaled minutes of each type of interaction increases. Panel C shows the change in odds ratio as the scaled minutes of each type of interaction increases. } 
    \label{fig:interactions}
\end{figure}

To uncover the interaction mechanisms of collaboration beyond pairs of participants, we focus on collaboration groups involving three participants.
In addition, we solely consider the first occurrence of a multi-year conference series to minimize the effect of prior knowledge in group formation.
We aggregate the data across all four conferences which corresponds to $52$ groups of three collaborating out of $113908$ total possible groups which did not.
We examine the impact of the four types of interaction defined above on team formation in groups of three.
As shown in Fig.~\ref{fig:interactions}, collaborators interacted more than non-collaborators for all four types of interaction. 

To further test the relationship of group interaction with collaboration, we fit the interaction data (in scaled minutes) to logistic models with linear combinations of the four types of interaction: synchronous interaction \textit{(sync1)}, asynchronous clique \textit{(async3)}, asynchronous wedge \textit{(async2)}, and asynchronous link \textit{(async1)}. We used stepwise selection to identify the best model fit by minimizing the Akaike Information Criterion (AIC) \cite{akaike1974} to identify which of the interaction types should be retained in the model to predict the group outcome \textit{(collaborated: 0/1)}. After evaluating all possible linear subcombinations of the variables, the stepwise selection retained all four, indicating that each significantly improved the AIC relative to models with fewer parameters. The final model took the form: 

\begin{equation}
    logit({collaborated}=1) = \beta_0+\beta_1 sync_1+\beta_2 async_3+\beta_3 async_2+\beta_4 async_1
\end{equation}
where the $\beta_i$ coefficients are the model parameters. Model results are displayed in Table~\ref{tab:stepwise-logistic}. All four covariates exhibited strongly positive coefficients and highly significant $p$-values ($10^{-4}$ or smaller), reflecting substantial increases in log-odds (and hence probability) of collaboration. The intercept of $-9.9$ implies a low baseline probability when all interaction types are zero, entailing that collaborations are unlikely to occur without one or more types of interaction. Compared to the intercept-only model, a likelihood ratio test produced a $\chi^2$ statistic of $127$ ($p=2.12\times 10^{-26}$), further confirming that including these interaction types is statistically justified. The odds ratios are calculated for a minute of scaled interaction, where ten minutes corresponds to approximately $30$ minutes of interaction in a group of three or $1.5$ hours in a group of nine. The increase in odds as interaction increases is shown in Panel C of Fig.~\ref{fig:interactions}. For example, in the case of synchronous interaction, an increase in $10$ minutes of scaled interaction increases the odds of collaboration $22$ times, whereas in the case of an asynchronous link, $10$ minutes increased the odds by $2.4$. The difference between the effect of the types of interaction is particularly visible in Panel B of Fig.~\ref{fig:interactions}, which shows how the probability of collaborating changes with increasing time. 

\begin{table}[htbp] 
\centering 
\begin{tabular}{lcccccc} 
\textbf{Term} & \textbf{Estimate ($\beta$)} & \textbf{SE} & \textbf{t-Stat} & \textbf{p-Value} & \textbf{Odds ratio} \\ 
\hline\\
Intercept & -9.9 & 0.36 & -28 & $5.4\times 10^{-167}$ & - \\ 
$\textrm{sync}_1$ & 0.31 & 0.025 & 12 & $2.3\times 10^{-34}$ & $1.4$\\ 
$\textrm{async}_3$ & 0.26 & 0.041 & 6.4 & $1.3\times 10^{-10}$ & $1.3$ \\ 
$\textrm{async}_2$ & 0.19 & 0.029 & 6.6 & $4.2\times 10^{-11}$ & $1.2$ \\ 
$\textrm{async}_1$ & 0.086 & 0.023 & 3.7 & $2.5\times 10^{-4}$ & $1.1$ \\ 
\\
\hline\\
\end{tabular} 
\caption{Final stepwise logistic regression model for interaction types and collaboration. The values presented are computed for scaled minutes (i.e. the odds ratio corresponds to the increase in odds for the increase in one scaled minute of interaction).} 
\label{tab:stepwise-logistic} 
\end{table}

\section{Discussion and Conclusion}

In this Chapter, we presented a framework for predicting team formation using higher-order network analysis.
This higher-order perspective will be useful for higher-order link prediction, using historical interaction and collaboration data to predict future collaborations.
Furthermore, our method for decomposing groups into different mechanisms provides a valuable contribution to our understanding of how network mechanisms such as triadic closure~\cite{simmel1908soziologie} and higher-order dynamics~\cite{iacopini2024temporal} predict future group interactions.
Lastly, one can apply our higher-order analysis to much larger scales such as bibliometric data~\cite{shi2023surprising} or research grant personnel~\cite{kardes2014complex,nakajima2023higher}.

The evidence presented in this chapter indicates that groups of three who collaborated interacted more than those who did not collaborate. The results suggest that both synchronous and asynchronous interaction, including in subgroups, play a role for team formation. While these results do not provide causal evidence, previous work studying these data from a pairwise perspective used quasi-random counterfactual conference schedule data to provide strong evidence for a causal link between conference interaction and team formation \cite{zajdela2025,Zajdela2022} and a similar analysis could be conducted for this higher-order network analysis. Furthermore, although the model presented here is a simple linear combination of the scaled interaction decomposed into four types, prior research~\cite{Zajdela2022} has shown that a nonlinear model that incorporates the effects of memory and captures the time dimension of interaction outperforms linear models. Future research could extend this nonlinear model to the higher-order interaction case for better predictive power. 

This research may have implications for designing conferences or other types of events aimed at fostering collaborative teams. The model results indicate that the probability of collaborating increases rapidly as a group interacts synchronously, particularly when the level of interaction is sustained over time and/or when the group is small. Given the large effect size of synchronous interaction on collaboration, this framework has implications when designing interactions. For example, conference organizers may want to consider not only the properties of pairs of participants assigned to interact in groups, but the properties of the larger-sized sub-groups as well. A question that remains is whether there is a saturation point at which the participants do not benefit from additional interaction time. Furthermore, in the digital age, interaction and collaboration in science have increasingly been conducted virtually, with tradeoffs for creativity and innovation \cite{brucks2022,lin2023}, team formation and community building \cite{zajdela2025}, environmental \cite{tao2021}, and equity considerations \cite{staples2006,wu2022,skiles2022}. This chapter focuses on in-person interaction, but we could apply a similar framework to virtual interaction, decomposed into its synchronous and asynchronous components.  

Lastly, we solely considered groups of three at scientific conferences in this chapter, but this formalization of the definition of interaction over time could be extended to larger groups in a variety of contexts. The framework could be applied to study the effect of interaction on other types of group outcomes, including the spread of innovations, behavior adoption, and opinion dynamics. It could also be applied to collective behavior observed in animal groups, for example in contexts where the decision to join or not join a group, mimic others, or learn a behavior may depend on time-varying interaction in higher-order groups. Thus, our approach demonstrates how higher-order network analysis can provide new insights on the dynamics of collaboration and team formation, with potential applications to collective behavior and collective intelligence.

\section*{Data availability}
A repository for this project, containing code and data to reproduce all results, is available on \href{https://github.com/nwlandry/higher-order-team-formation}{GitHub} and at Ref.~\cite{software}.\\

\acknowledgments{The authors thank the Research Corporation for Science Advancement for providing the Scialog dataset, Sandeep Chowdhary for initial conversations, and Simon Levin for discussions about defining group interaction. E.R.Z. acknowledges Elena Graetz and Maher Said for their input on the statistical methods used. N.W.L. acknowledges support from the University of Virginia Prominence-to-Preeminence (P2PE) STEM Targeted Initiatives Fund, SIF176A Contagion Science. This research was supported by an appointment to the Intelligence Community Postdoctoral Research Fellowship Program at Princeton University administered by Oak Ridge Institute for Science and Education (ORISE) through an interagency agreement between the U.S. Department of Energy and the Office of the Director of National Intelligence (ODNI). The work in this chapter was supported by the Siegel Research Fellowship and a gift from William H. Miller III.}

\bibliography{bibliography}

\end{document}